# STUDIES OF COUPLED CAVITY LINAC (CCL) ACCELERATING STRUCTURES WITH 3-D CODES[*]

G. Spalek, D.W. Christiansen, P.D.Smith, P.T. Greninger, C.M. Charman, General Atomics, San Diego, CA 92186, USA

R.A. DeHaven, Techsource, Inc., Santa Fe, NM 87501, USA

L.S.Walling, Ansoft Corp., Albuquerque, NM 87113

*Abstract*

The cw CCL being designed for the Accelerator Production of Tritium (APT) project accelerates protons from 96MeV to 211MeV. It consists of 99 segments each containing up to seven accelerating cavities. Segments are coupled by intersegment coupling cavities and grouped into supermodules. The design method needs to address not only basic cavity sizing for a given coupling and pi/2 mode frequency, but also the effects of high power densities on the cavity frequency, mechanical stresses, and the structure's stop band during operation. On the APT project, 3-D RF (Ansoft Corp.'s HFSS) and coupled RF/structural (Ansys Inc.'s ANSYS) codes are being used to develop tools to address the above issues and guide cooling channel design. The code's predictions are being checked against available low power Aluminum models. Stop band behavior under power will be checked once the tools are extended to CCDTL structures that have been tested at high power. A summary of calculations made to date and agreement with measured results will be presented.

## 1 INTRODUCTION

To design a CCL structure, the RF design codes need to predict the pi/2 mode frequency ($f_{pi/2}$) and nearest (k) and next-nearest-neighbor coupling constant (kk) of the CCL. Calculation accuracy has to be such that final frequency adjustment of fabricated structures can be accomplished by the usual methods of machining of reasonably sized tuning features (rings) or by small mechanical deformation of thinned portions of cavity walls. The pi/2 mode frequency is arrived at by calculating the frequency of the accelerating cavities (ac) without coupling slots and correcting this frequency by frequency shifts caused by coupling slots and next-nearest-neighbor coupling [1]. To obtain as high an accuracy as possible, the approach used in this work was to calculate the no slot absolute frequencies in 2-D using SUPERFISH and only calculate frequency differences in 3-D using HFSS.

High RF power operation of the accelerator causes thermally induced stresses that can lead to:
- Yielding of the cavity metal components leading to permanent frequency shifts.
- Excessive frequency shifts of the accelerating mode.
- Different frequency shifts for the coupling and accelerating cavities (i.e. change in stop band width leading to lower field stability).

To predict high power effects, a code like ANSYS that couples electromagnetic, thermal, and structural calculations can be used to calculate the mechanical distortions and stresses caused by a particular cooling channel configuration and power level. The frequencies of the accelerating and coupling modes of the distorted structure can then be calculated in two ways:
- By recalculating the modes of the distorted structures using the deformed geometry ("morphed" mesh), the coupled thermal-mechanical problem.
- By calculating the frequency shifts caused by the cavity deformations using a technique such as the Slater [2] perturbation method.

Since the expected frequency shifts are small (e.g. 100kHz for a 700MHz cavity), the perturbation method could be expected to yield more accurate results than the recalculation of the entire problem with a deformed and re-meshed geometry.

This article reports on results of the HFSS and ANSYS calculations performed to date.

## 2 RF CALCULATIONS_RESULTS

### 2.1 Calculation Procedure

The dimensions of a half scale (~1,400MHz) Aluminum model of segment 243 of the APT CCL were used for the calculations. The calculation procedure was:
- Calculate the frequency $f_{0sf}$ of the accelerating cavity using SUPERFISH [3].
- Construct an HFSS model of one period of the CCL segment consisting of ½ accelerating cavity and two ¼ sections of coupling cavities.

---

[*] Work supported by U.S. DOE contract DE-AC04-96AL89607

- Use periodic boundary conditions with coupling cavity field phase shifts of $0^0$, $36^0$, $72^0$, $108^0$, $144^0$, and $180^0$ to calculate the mode spectrum of the structure with HFSS.
- Fit the mode spectrum to obtain the cavity frequencies ($f_{hfss}$) including slots, and $k_{hfss}$ and $kk_{hfss}$.
- Generate a model of ½ of the accelerating cavity in HFSS <u>using the same segmentation</u> as in the periodic model (to guide the mesh and control cavity effective size ) and calculate its frequency $f_{0hfss}$.
- Calculate the (HFSS) accelerating cavity frequency shift

$$df_{hfss} = f_{hfss} - f_{ohfss}$$

- Calculate the pi/2 mode frequency:

$$f_{pi/2hfss} = \frac{f_{0sf} + df_{hfss}}{\sqrt{1 - kk_{hfss}}}$$

Calculations were performed for several cases that included knife edged coupling slots with different coupling cavity depth (case 1 and case 2 below) and the enlargement of coupling slots by chamfering (case 3 and case 4 below). Frequency shifts were calculated for ac cavities only, future work will address the coupling cavities. For comparison with the calculations, Aluminum model no-slot ac frequencies and mode frequencies were measured. The mode spectra of the cavity model (11 cavities) were fit to extract the ac frequencies and coupling constants.

## 2.2 Results/Comparison With Measurements

Table 1 compares the calculated values and quantities extracted from fits of calculated and measured mode spectra.

Table 1: Comparison with measured values, frequencies are in MHz

|  | Case 1 | Case 2 | Case 3 | Case 4 |
|---|---|---|---|---|
| $f_{0sf}$ | 1,432.352 | 1,432.352 | 1,432.352 | 1,432.352 |
| $f_{0hfss}$ | 1,436.960 | 1,436.960 | 1,436.960 | 1,436.960 |
| $f_{hfss}$ | 1,419.489 | 1,412.12 | 1,414.235 | 1,412.989 |
| $df_{hfss}$ | -17.471 | -19.047 | -22.725 | -23.972 |
|  |  |  |  |  |
| $f_{pi/2hfss}$ | 1,411.209 | 1,409.541 | 1,404.776 | 1,403.117 |
| $f_{pi/2mea}$ | 1,409.886 | 1,408.318 | 1,403.871 | 1,401.608 |
| diff | 1.32 | 1.22 | 0.91 | 1.5 |
|  |  |  |  |  |
| $k_{hfss}$ | 0.03712 | 0.03981 | 0.04489 | 0.04671 |
| $k_{meas}$ | 0.03764 | 0.04005 | 0.0444 | 0.0464 |
| dif(%) | 1.4 | 0.6 | -1.1 | 0.67 |
|  |  |  |  |  |
| $kk_{hfss}$ | -0.00521 | -0.00535 | -0.00692 | -0.00752 |
| $kk_{meas}$ | -0.00556 | -0.00554 | -0.00705 | -0.00785 |
| dif(%) | 6.7 | 3.5 | 1.9 | 4.4 |

The differences between the calculated and measured pi/2 mode frequencies are well within the correction range (+/-3MHz) for a reasonably sized tuning ring. The nearest-neighbor coupling constants are very accurately calculated and the accuracy of the next-nearest-neighbor coupling calculation is also reasonable.

## 2 ANSYS CALCULATIONS

Post-processor ANSYS routines implementing the Slater perturbation calculations in 3-D have been written. To test the implementation, frequency shifts for simple geometry cavities were calculated and checked by comparison to theory and to the fully coupled calculations. The calculated cases included:

- Copper 1GHz pillbox cavity ($TM_{010}$ mode), frequency shift due to a uniform $-10^0F$ temperature change.
- Copper pillbox cavity, outside surface of metal held at constant temperature, inside heated by radio frequency (RF) fields.
- Copper CCL cavity with cooling channels heated by RF fields.

In addition, work is proceeding on modelling and calculations of modes and frequency shifts for a periodic APT CCL structure.

### 3.1 Pillbox Cavity Results

The radius of a 1GHz pillbox is:

r=11.4742541cm

Using an expansion coefficient for copper of $10^{-5}$ /$^0$F, a 10 degree lowering of temperature gives a new frequency:

$$f = \frac{r_{01}c}{2p(r+dr)} = 1,000,1000,011 Hz$$

Where $r_{o1}$ is the root of $J_0$, c is the speed of light, and dr is the change in radius.
So the frequency shift is:

$df_{calc}$=100,011Hz

Using the same radius change and the surface fields calculated by ANSYS, the Slater perturbation method gives a frequency shift of:

$df_{slater}$=100,083Hz

The corresponding calculation results of the absolute cavity frequency using ANSYS linear and quadratic basis functions are:

$f_{linear}$=1,001,028,656Hz

$f_{quadratic}=1{,}000{,}687{,}772$ Hz

The error in calculating the absolute frequency of the pillbox is larger (~688kHz) than the frequency shifts of interest in the design of accelerator cavities. This error is due to mesh size, numerical accuracy, shape function order, etc. This indicates that errors can be expected to grow as the geometry of the cavity becomes more complicated by coupling slots, noses, and other features that increase the curvature of surfaces, distort the mesh, etc.

Coupled calculations of a Copper pillbox were used to calculate the frequency shifts of a thick wall cavity heated internally by RF fields while the outside of the Copper was held at constant temperature. Two cases were calculated, one without including the effects of the motion of the cavity circular end walls and the other including these effects. The motion of the end walls should result in zero frequency shift and is thus a good test of the accuracy of calculation of electric and magnetic contributions to the frequency shift in the Slater method.

The results are compared to the Slater calculations and theory below.

The heating-caused radius change calculated by ANSYS was $4.8551*10^{-2}$cm. This gives a calculated frequency shift of :

$df_{calc}=-4{,}213{,}469$Hz

The corresponding results of the coupled and Slater calculations using quadratic shape functions for the end-wall and no-end wall cases are listed in Table 2.

Table2: Comparison of Copper Pillbox Results

|  | w/o end wall | w/ end wall |
|---|---|---|
| $df_{coupled}$ | -4,248,333Hz | -4,181,683Hz |
| $df_{slater}$ | -4,234,887Hz | -4,232,266Hz |
| $df_{calc}$ | -4,213,469Hz | -4,213,469Hz |

The 0.06% difference in the Slater end-wall/no-end-wall results shows the degree of non-cancellation of the electric and magnetic field contributions to the frequency shifts on the end walls of the cavity. This error is quite small and the Slater agreement with the calculated value is excellent (0.4%). The larger 1.5% difference in the coupled results is probably due to the morphing of the mesh near the end-walls of the cavity.

### 3.2 CCL Cavity Results

Only preliminary calculations were made for the axi-symmetric CCL cavity with cooling channels.

For a uniform $-10^0$F change in the temperature of a Copper CCL Cavity the resultant frequency shift results are:

$df_{calc}=71{,}299$Hz

$df_{slater}=73{,}100$Hz

The above 2.5% difference is larger than expected and needs to be investigated to make sure that displacements are calculated correctly in areas of small radii of curvature.

The results for the cavity cooled by cooling channels are:

$df_{coupled}=-198{,}202$Hz

$df_{slater}=-202{,}368$Hz

The above frequency shifts differ by only 2.1%. They will be compared to 2-D calculations in the future.

## 2 CONCLUSIONS

The comparisons made above indicate that:
- 3-D codes such as HFSS can be used to calculate coupling constants of periodic CCL structures very accurately.
- Frequency shifts caused by coupling slots can be calculated by 3-D codes and used to predict pi/2 mode frequencies with sufficient accuracy to be useful for design.

If further comparisons between calculations and experiment are as successful as those above, 3-D accelerator structures can be designed with 3-D codes without the construction and testing of numerous Aluminum models.

The preliminary work above also indicates that a 3-D Slater perturbation method can be used to predict thermally caused frequency shifts in accelerator structures while avoiding possible problems associated with mesh morphing in the ANSYS code. This method also shortens calculation times since a recalculation of the resonant mode frequencies is not necessary after the thermal/mechanical calculations are completed.